\documentclass{PoS}

\usepackage{multirow}

\newcommand{\CK}{\v Cerenkov}

\title{Astrophysics with the AMS-02 experiment}

\ShortTitle{Astrophysics with the AMS-02 experiment}

\author{\speaker{Rui Pereira}\\
        on behalf of the AMS collaboration\\
        LIP Lisbon\\
        Lisbon, Portugal\\
        E-mail: \email{pereira@lip.pt}}

\abstract{
The Alpha Magnetic Spectrometer (AMS), whose final version AMS-02 is to be
installed on the International Space Station (ISS) for at least 3 years, is
a detector designed to measure charged cosmic ray spectra with energies up
to the TeV region and with high energy photon detection capability up to a
few hundred GeV, using state-of-the-art particle identification techniques.
Following the successful flight of the detector prototype (AMS-01) aboard
the space shuttle, AMS-02 is expected to provide a significant improvement
on the current knowledge of the elemental and isotopic composition of
hadronic cosmic rays due to its long exposure time (minimum of 3 years) and
large acceptance (0.5 m$^2$sr) which will enable it to collect a total
statistics of more than $10^{10}$ nuclei. Detector capabilities for charge,
velocity and mass identification, estimated from ion beam tests and
detailed Monte Carlo simulations, are presented. Relevant issues in cosmic
ray astrophysics addressed by AMS-02, including the test of cosmic ray
propagation models, galactic confinement times and the influence of solar
cycles on the local cosmic ray flux, are briefly discussed.
}

\FullConference{International Europhysics Conference on High Energy Physics\\
                 July 21st - 27th 2005\\
                 Lisboa, Portugal}

\begin{document}

\section{The AMS experiment}

Alpha Magnetic Spectrometer (AMS)\cite{bib:ams} is an experiment designed
to study the cosmic ray flux by direct detection of particles above the
Earth's atmosphere. The the final detector (AMS-02) will be ready for
launch by the end of 2007 and installed on the International Space Station
(ISS) for more than 3 years. A preliminary version of the detector (AMS-01)
was successfully flown aboard the US space shuttle Discovery in June 1998.

\begin{figure}
\vspace{-0.7cm}
\begin{minipage}[b]{8.3cm}
\center
\includegraphics[width=1.05\textwidth]{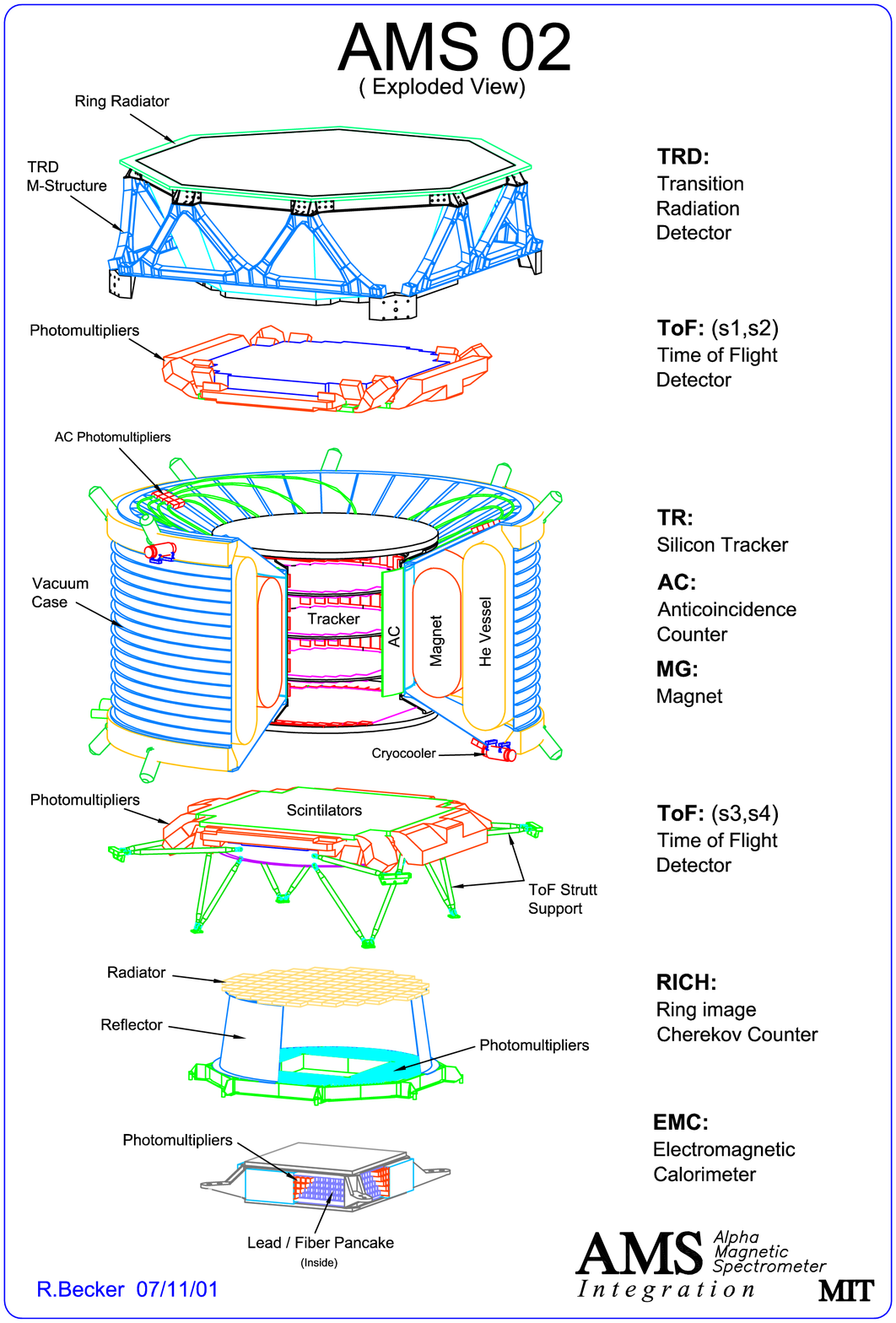}
\caption{Exploded view of the AMS-02 detector}
\label{amsdet}
\end{minipage}
\hfill
\begin{minipage}[b]{6.0cm}
\center
\vspace{0.6cm}
\hspace{-0.5cm}
\includegraphics[width=0.9\textwidth]{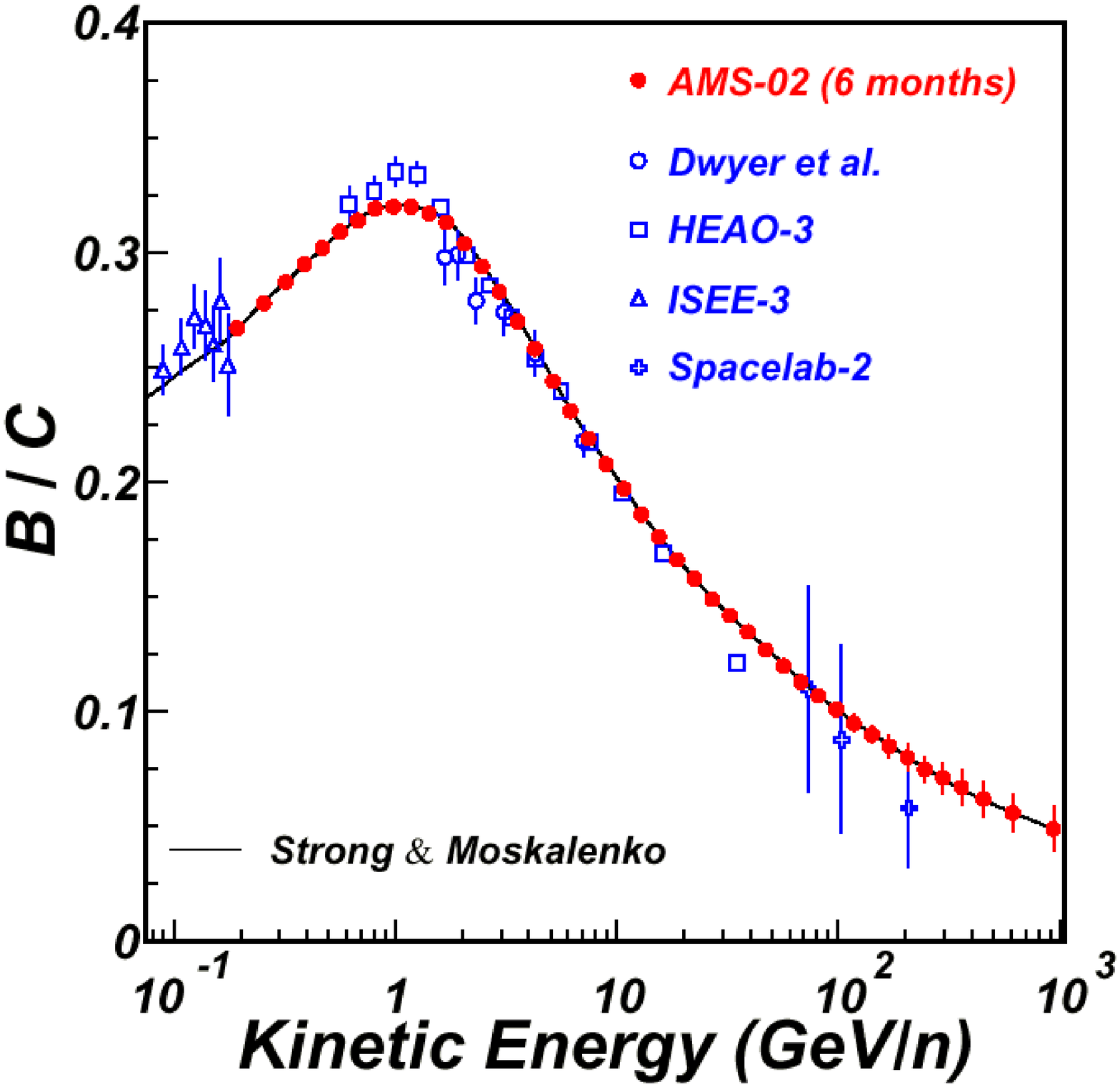}
\\
\vspace{-0.2cm}
\caption{AMS-02 expected performance for elemental ratio B/C after 6 months
of data taking compared with existing results.}
\label{amsprosp2}
\includegraphics[width=1.0\textwidth]{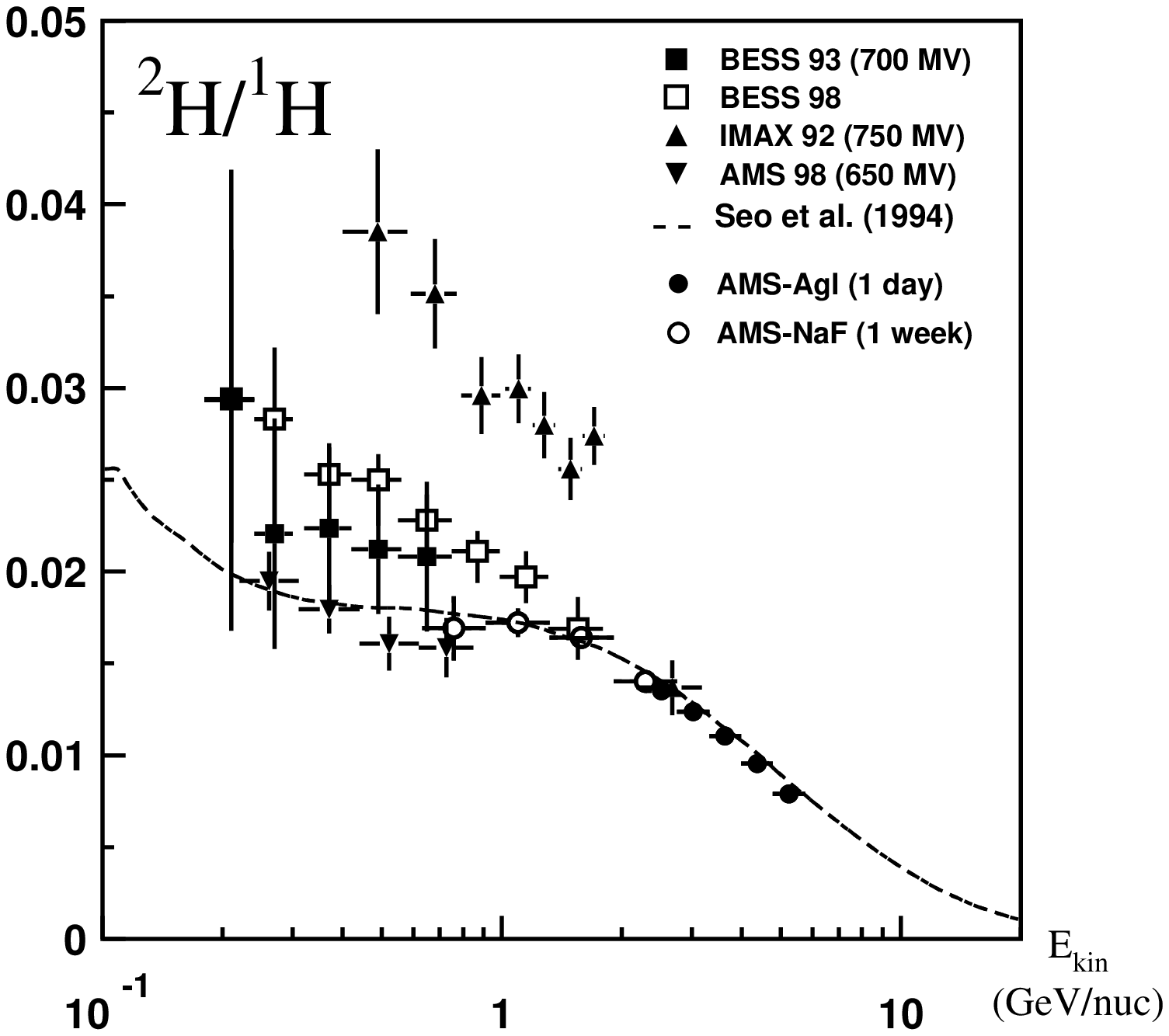}
\\
\vspace{-0.5cm}
\caption{Reconstruction of simulated isotopic ratio D/p in AMS-02 compared
with results from previous experiments.}
\label{amsprosp1}
\end{minipage}
\hfill
\end{figure}

On the ISS, orbiting at an average altitude of 400 km, AMS-02 will collect
an extremely large number of cosmic ray particles. Its main goals are
\emph{(i)} a detailed study of cosmic ray composition and energy spectrum
through the collection of an unprecedented volume of data, \emph{(ii)} a
search for heavy antinuclei (\mbox{$Z \geq$ 2}) which if discovered would
signal the existence of cosmological antimatter (the detection of
anti-carbon (\mbox{$Z=6$}) would be evidence for the existence of
anti-stars), and \emph{(iii)} a search for dark matter constituents by
examining possible signatures of their presence in the cosmic ray spectrum.
AMS-02 is equipped with a superconducting magnet cooled by superfluid
helium. The spectrometer is composed of several subdetectors: a Transition
Radiation Detector (TRD), a Time-of-Flight (TOF) detector, a Silicon
Tracker, Anticoincidence Counters (ACC), a Ring Imaging \CK\ (RICH)
detector with a dual radiator (silica aerogel and sodium fluoride) and
an Electromagnetic Calorimeter (ECAL). Fig.~\ref{amsdet} shows a schematic
view of the full AMS-02 detector.

\section{Detector requirements}

To fulfill its astrophysics goals, AMS-02 must be able to detect nuclei and
identify their charge in a large range of $Z$. Precise velocity and
rigidity measurements are also required, namely to discriminate between
different isotopes of the same element.

Cosmic ray propagation models may be tested by comparing abundances of
primary and secondary species at different energies. Primary cosmic rays
are those produced at the original sources, while secondaries are created
through the interaction of primary cosmic rays with the interstellar
medium. Abundance ratios such as the one between boron ($Z=5$, secondary)
and carbon ($Z=6$, primary), shown in fig.~\ref{amsprosp2}, depend on the
details of cosmic ray propagation.

Data on confinement times may be inferred from the comparison between
stable and unstable nuclei with similar origin. This is the case of
beryllium isotopes ${}^9 \textrm{Be}$ (stable) and ${}^{10} \textrm{Be}$
(radioactive, $t_{1/2} = \textrm{1.6~Myr}$), which are both secondary
species.

\section{Measurement of particle properties}

In AMS-02 the evaluation of particle charge is performed using data from
the Silicon Tracker, TOF and RICH detectors. In the Tracker and TOF, charge
is given by direct sampling of local energy deposition along the particle's
path: $\Delta E \propto Z^2$. In the case of the RICH detector the energy
deposition is also proportional to $Z^2$ but it is spread over a large
region (\CK\ ring). The charge sign is determined from the bending of the
trajectory in the Silicon Tracker, combined with albedo rejection
capabilities from the TOF and RICH detectors. The expected charge
resolution $\Delta Z$ from redundant measurements in the TOF, Tracker and
RICH is $\Delta Z \lesssim 0.25$ charge units up to the iron region ($Z=26$).

Particle velocity may be obtained from the TOF and RICH detectors. In the
Time of Flight detector, velocity is calculated from the crossing time
$t$ between scintillator planes. In AMS-02 the expected accuracy for
protons is $\Delta t \sim \textrm{130~ps}$, leading to
$\Delta \beta / \beta \sim 4 \%$. In the RICH detector, velocity is
calculated from the aperture of the \CK\ cone ($\theta_c$):
$cos~\theta_c = \frac{1}{\beta n}$ where $n$ is the radiator's refractive
index. The typical accuracy in AMS-02 is
$\Delta \beta / \beta \sim 10^{-3}$ for $Z=1$, and
$\Delta \beta / \beta \sim 10^{-4}$ for $Z>10$. However, the RICH
measurement is only available if the particle's kinetic energy $E_{kin}$
exceeds the \CK\ threshold for the radiator crossed:
$E_{kin} > \textrm{0.5~GeV/nucleon}$ for sodium fluoride (NaF) ($\sim 10\%$
of events), and $E_{kin} > \textrm{2.1~GeV/nucleon}$ for silica aerogel
(the remaining $\sim 90\%$).

Additionally, the Transition Radiation Detector combined with the
Electromagnetic Calorimeter provides $e/p$ separation up to
$\sim \textrm{1~TeV}$ with a discrimination factor of $\sim 10^5$.

Mass identification in AMS-02 relies on precise velocity and rigidity
measurements, where the rigidity $R$, defined as $R = p/Z$, is provided
by the sampling of the particle's trajectory given by the Silicon Tracker.
For protons with $E_{kin} \sim \textrm{1-10~GeV}$, the accuracy is
$\Delta R/R \sim 2 \%$. The final mass estimate is given by
$m = \frac{RZ}{\gamma \, v}$ where $\gamma$ is the Lorentz factor. The
expected mass resolution is of the order of $\textrm{2-3\%}$ for light
nuclei ($Z \lesssim 5$) with $E_{kin} \sim \textrm{5~GeV/nucleon}$. The
effect of the mass resolution in the isotopic separation depends also on
the isotopic ratio. Separation of isotopes will be possible up to high
kinetic energies: at least 6~GeV/nucleon for hydrogen
(fig.~\ref{amsprosp1}), where the limit comes from the large proton
background ($D/p \sim 10^{-2}$), and 8~GeV/nucleon in the case of elements
such as He and Be.

\section{AMS-02 prospects}

Fig.~\ref{amsprosp2} shows the expected performance in the determination of
the elemental ratio B/C after 6 months of data taking. AMS-02 results will
represent a major improvement with respect to existing data, giving
information on abundances of these (and other) elements up to energies of
$\sim$~1~TeV/nucleon.

Fig.~\ref{amsprosp1} shows the results for the isotopic ratio D/p as a
function of energy in two simulations of hydrogen events\cite{bib:isotsep},
one consisting only of particles crossing the NaF radiator of the RICH
detector and the other corresponding to particles crossing the aerogel
radiator. Events were simulated according to the reacceleration model of
reference \cite{bib:seo:h} represented by the dashed line. The total
statistics was $\sim 1.5 \times 10^7$ events in each case, which
corresponds to $\sim$~1~week of data in the case of NaF and $\sim$~1~day in
the case of aerogel. In both cases, the results obtained show a large
improvement with respect to existing data from previous experiments.

It should be noted that solar cycles have a significant effect on the ratio
D/p (effects on heavier elements are much smaller). The long operating
period of AMS-02 will allow to collect data over a significant fraction of
a solar cycle (11 years on average between successive maxima), and its high
acceptance will provide enough statistics to obtain ratio samplings every
few days.

\section{Conclusions}

Data from AMS-02 will provide a major improvement with respect to existing
results for the hadronic cosmic ray spectrum. A total statistics of more
than $10^{10}$ events will be collected during its operation. Detector
capabilities include charge separation up to $Z \sim 30$, velocity
reconstruction with $\Delta \beta / \beta \sim 10^{-3}$ for $Z=1$ and
$\Delta \beta / \beta \sim 10^{-4}$ for $Z \sim 10-20$, and isotopic
separation of light elements. Results of AMS-02 will address key issues in
cosmic ray astrophysics, namely, propagation of cosmic rays in the Galaxy,
confinement times, and the effect of solar cycles on the flux in Earth's
vicinity.

\end{document}